\documentclass[a4paper,12pt]{article}
\usepackage{a4}
\usepackage{graphics}
\usepackage{epsfig}
\usepackage{amssymb}
\usepackage{color}

\definecolor{green2}{rgb}{0.1,0.9,0.0}
\definecolor{green3}{rgb}{0.,0.7,0.0}
\definecolor{green4}{rgb}{0.,0.8,0.4}
\definecolor{lightblue}{rgb}{0.0,1.0,1}
\definecolor{blue00}{rgb}{0.,0.1,1}
\definecolor{blue0}{rgb}{0.0,0.0,0.5}
\definecolor{blue2}{rgb}{0.1,0.1,0.4}
\definecolor{blue3}{rgb}{0.,0.4,0.8}
\definecolor{red1}{rgb}{0.9,0,0}

\newcommand{\la}{\lambda}

\newcommand{\eps}{\epsilon}

\newcommand{\neu}{\tilde{\chi}^0}
\newcommand{\neutralino}{\tilde{\chi}^0}
\newcommand{\slepton}{\tilde{\ell}}
\newcommand{\stau}{\tilde{\tau}}

\newcommand{\sleptonmp}{\tilde{\ell}^\mp}

\newcommand{\etarl}{\eta_{\ell}^n}

\newcommand{\etatauk}{\eta_{\tau}^n}
\newcommand{\etatau}{\eta_{\tau}^1}

\begin{document}
\hfill WUE-ITP-2004-014\\
\hspace*{\fill} hep-ph/0407057\\
\begin{center}
{\Large \bf {Interference of Higgs boson resonances in $\mu^+\mu^-\to\neutralino_i\neutralino_j$ with longitudinal beam polarization}}
\end{center}
\vspace{0.5cm}
\begin{center}
{\large \sc		
H.~Fraas, F.~von~der~Pahlen, C.~Sachse}
\end{center}
\begin{center}
{\small \it  	Institut f\"ur Theoretische Physik, Universit\"at W\"urzburg, Am Hubland, \\
		D-97074 W\"urzburg, Germany
}
\end{center}
\begin{abstract}
We study 
the interference of resonant Higgs boson exchange
in neutralino production in $\mu^+\mu^-$ annihilation with longitudinally polarized beams. 
We use the energy distribution of the decay lepton in the process
$\neutralino_j \to \ell^{\pm} \sleptonmp$ to determine the polarization of the neutralinos. 
In the CP conserving Minimal Supersymmetric Standard Model
a non-vanishing asymmetry in the lepton energy spectrum 
is caused by
the interference of Higgs boson exchange channels with different CP eigenvalues.
The contribution of this interference is large if the heavy neutral bosons $H$ and $A$ are nearly degenerate.
We show that the asymmetry 
can be used to
determine the couplings of the neutral Higgs bosons
to the neutralinos.
In particular, the asymmetry allows to determine the relative phase of the couplings.
We find large asymmetries and cross sections for a set of reference scenarios
with nearly degenerate neutral Higgs bosons.
\end{abstract}
\newpage
\section{Introduction}
At a muon collider, neutral Higgs bosons are produced as s-channel resonances in $\mu^+\mu^-$ annihilation \cite{hefreports,mucolhiggs,barger1}.
Therefore a muon collider is an excellent tool to study the properties of a heavy scalar or pseudoscalar neutral Higgs boson.
The CP conserving Minimal Supersymmetric Standard Model (MSSM) contains three neutral Higgs bosons, 
a light scalar $h$, a heavier scalar $H$, and a pseudoscalar $A$. 
Since the two heavier neutral Higgs bosons may decay into neutralinos 
a muon collider opens the possibility to test supersymmetry
through the interaction of the Higgs sector with the neutralino sector. 

Neutralinos, the supersymmetric partners of the neutral Higgs and gauge bosons, are expected to be light in large regions of the supersymmetric parameter space.
A scan of their production line shape may allow to separate the contributions from the different resonances \cite{barger1} and a determination of
the coupling mechanism of the Higgs sector to the higgsino and gaugino sectors \cite{higgsneutralino}. 
However, 
the two heavier resonances are nearly degenerate over much of the ($\tan\beta,m_A$) parameter space and a large overlap will make the determination of the resonance parameters a difficult task.
The production process dependence 
on the polarizations of the initial muons and that of the neutralinos 
may allow to disentangle the contributions of the different resonances.

In this paper we study the $\mu^+\mu^-\to \neu_i\neu_j$ production process with longitudinal muon beam polarization, where $i$ and $j$ label the mass eigenstates of the neutralinos.
The interference of the Higgs boson exchange channels with different CP quantum numbers can be sizable 
when the mass difference of these Higgs bosons is at most of the same order of their decay widths \cite{asakawa}.
The neutralino polarization, averaged over the neutralino production angles,
results from 
the interference of the neutral Higgs bosons
since
the contributions from $Z$ and $\tilde{\mu}$ exchange 
vanish, due to the Majorana character of the neutralinos \cite{gudi.majorana}.
In the center of mass system (CMS) 
the energy distribution 
of the decay lepton of the decay $\neu_j\to\ell\slepton$
depends on the longitudinal neutralino polarization. 
The interference effect is proportional to the sum of the polarizations of the initial 
fermions, as well as that of the neutralinos,
and thus vanishes for unpolarized beams. 

This paper is organized as follows: in Section~2 we present the formalism
and discuss the lepton energy distribution asymmetry. 
In Section~3 we study the possibility of using the asymmetry to determine the Higgs neutralino couplings.
In Section~4 we present numerical results and 
in Section~5 we present a short summary and draw the conclusions.


\section{Definitions and Formalism}

We study in the MSSM neutralino production 				
\begin{equation}\label{prod}
\mu^+\ \mu^-\rightarrow\tilde{\chi}_i^0\ \tilde{\chi}_j^0
\label{eq:process}
\end{equation}
with longitudinally polarized beams for center of mass energies 	
around the resonances of the heavy neutral Higgs bosons $H$ and $A$,
and the subsequent leptonic two-body decay of one of the neutralinos
\begin{eqnarray}
\label{dec}
&&\tilde{\chi}_j^0\rightarrow\ell^\pm\tilde{\ell}^\mp,\quad \ell=e,\mu,\tau.
\end{eqnarray}

The production proceeds via the resonant exchange of $H$ and $A$ in the s-channel,
as well as via the non-resonant exchange of the 
$Z$ boson and of the light Higgs boson $h$ in the s-channel 
and of the t- and u-channel exchange of $\tilde{\mu}_{L,R}$.

\subsection{Lagrangian and couplings}
%
\label{section:lagrdens}
The interaction Lagrangians for neutralino production via Higgs exchange are (in our notation we follow closely 
\cite{HK,GH,gudi99})
\begin{eqnarray}
\label{mumuphi}
{\cal L}_{\mu^+ \mu^- \phi} & = &
          g \, 
                \bar{\mu}\, 
          (c^{(\phi\mu)\,*} P_L + c^{(\phi\mu)} P_R)
		\, \mu \, \phi,\\
{\cal L}_{\tilde{\chi}^{0}_i \tilde{\chi}^{0}_j \phi} & = &
          \frac{1}{2}g \, \bar{\tilde{\chi}}_i^0
          (c^{(\phi)\,*}_{ij} P_L + c^{(\phi)}_{ij} P_R)
          \tilde{\chi}_j^0\,\phi,
\label{chichiphi}
\end{eqnarray}
where $P_{R,L}=\frac{1}{2}(1\pm\gamma^5)$, $g$ is the weak coupling constant and $\phi=H,A,h$.
In the neutralino basis $\{\tilde{\gamma}, \tilde{Z}, H_1^0, H_2^0\}$ 
the muon and neutralino couplings to $H$ and $A$ are \cite{GH} 
\begin{eqnarray}
	c^{(H\mu)} &=& -\frac{m_\mu}{2 m_W}\frac{\cos\alpha}{\cos\beta},
\label{eq:cHmu}
\\	c^{(A\mu)} &=& i \frac{m_\mu}{2 m_W}\tan\beta,
\\
         c^{(H)}_{ij} &=& - Q_{ij}^{\prime\prime}\cos\alpha  +  S_{ij}^{\prime\prime}\sin\alpha,
\label{eq:cH}
\\         
	c^{(A)}_{ij} &=& -i(  Q_{ij}^{\prime\prime}\sin\beta -  S_{ij}^{\prime\prime}\cos\beta),
\label{eq:cA}
\\
	Q_{ij}^{\prime\prime} &=& \frac{1}{2\cos\theta_W}[N_{i3}N_{j2}+ (i\leftrightarrow j)],
\\
	S_{ij}^{\prime\prime} &=& \frac{1}{2\cos\theta_W}[N_{i4}N_{j2}+ (i\leftrightarrow j)],
\end{eqnarray}
where $\alpha$ is the Higgs mixing angle, $\tan\beta=v_2/v_1$ is the ratio of the vacuum expectation values of the two neutral Higgs fields,
 $\theta_W$ is the weak mixing angle
and $N$ is the unitary $4\times 4$ matrix which diagonalizes the neutralino mass matrix $Y$.
If $CP$ is conserved $Y$ is real and the matrix $N$ can be chosen real and orthogonal: 
$N_{i\alpha}Y_{\alpha\beta}N^T_{\beta k}=\eta_i m_{\chi_i}\delta_{ik}$,
where $m_{\chi_i}, i=1,\ldots,4$ are the masses of the neutralinos and $\eta_i=\pm 1$ is related to the $CP$ eigenvalue of the neutralino $\neu_i$. 
	The muon and neutralino couplings to the lighter Higgs boson $h$ are obtained substituting $\alpha$ by $\alpha+\pi/2$ in
	eqs.~(\ref{eq:cHmu}) and (\ref{eq:cH}). 

The interaction Lagrangian for neutralino decay into a lepton and a slepton of the first two generations is given by
\begin{eqnarray}
	{\cal L}_{\ell \slepton \neu_j} & = & 
		g f_{\ell_j}^L \bar{\ell} P_R \neu_j  \slepton_L 
	+	g f_{\ell_j}^R \bar{\ell} P_L \neu_j  \slepton_R
	+	\mbox{h.c.},
\label{eq:Lagneutoellsel}
\end{eqnarray}
with couplings
\begin{eqnarray}
	f_{\ell j}^L &=& -\sqrt{2}\bigg[\frac{1}{\cos
	\theta_W}(T_{3\ell}-e_{\ell}\sin^2\theta_W)N_{j2}+
	e_{\ell}\sin \theta_W N_{j1}\bigg],
\label{eq:fl}\\
	f_{\ell j}^R &=& -\sqrt{2}e_{\ell} \sin \theta_W\Big[\tan 
	\theta_W N_{j2}^*-N_{j1}^*\Big],
\label{eq:fr}
\end{eqnarray}
where $e_{\ell}$ and $T_{3\ell}$ denote the electric charge and third component of the weak isospin of the lepton $\ell$.

Mixing can safely be neglected for the scalar leptons of the first two generations, $\slepton = \tilde{e},\tilde{\mu}$. 
For the neutralino decay into staus
$\neu_i \to  \stau_n \tau$, we take stau mixing into account and write for the Lagrangian \cite{Bartl:2002bh}
\begin{eqnarray}
	& & {\mathcal L}_{\tau \stau \chi_i }=  g \stau_n \bar \tau
	(a^{\stau}_{nj} P_R+b^{\stau}_{nj} P_L)\neu_i + {\rm h.c.}~,
	 \quad n = 1,2; \; j=1,\dots,4, 
\label{eq:Lagneutotaustau}
\end{eqnarray}
where the coefficients $a^{\stau}_{nj}$ and $b^{\stau}_{nj}$ are given in Appendix~\ref{tau.neutralino}.

\subsection{Cross section and lepton energy distribution}

To calculate the cross section for the combined process of neutralino production and decay 
we use the spin density matrix formalism of \cite{Haber94}, as e.g.\ for neutralino production in $e^+ e^-$ annihilation in \cite{gudi99}.

The spin density matrices $\rho^{P}$ of $\neu_i\neu_j$ production and $\rho^{D}$ of $\neu_j$ decay are given by
\begin{eqnarray}
	\rho^{P}_{\la_j\la^\prime_j} &=& \sum_{\la_i} 
		{T}^{P}_{\la_i\la_j}{{T}}^{P*}_{\la_i\la^\prime_j},
\label{rhop}
\\
	\rho^{D}_{\la^\prime_j\la_j} &=& 		
		{{T}}^{D\ast}_{\la^\prime_j}{{T}}^{D}_{\la_j},
\label{rhod}
\end{eqnarray} 
where $T^P_{\la_i\la_j}$ and $T^D_{\la_j}$ denote the helicity amplitudes for production and decay, respectively. 
The amplitude squared for production and decay is then
\begin{eqnarray}
	|{{T}}|^2 &=& |\Delta(\neu_j)|^2 \sum_{\la_j \la_j^\prime} \rho^P_{\la_j \la_j^\prime}\rho^D_{\la_j^\prime\la_j},
\label{tsquare}
\end{eqnarray}
with the propagator $\Delta(\neu_j)=i/[p^2_{\chi_j}-m_{\chi_j}^2+i m_{\chi_j}\Gamma_{\chi_j}]$.
Here 
	$p_{\chi_j}^2$, $m_{\chi_j}$ and $\Gamma_{\chi_j}$ denote the four-momentum squared, mass and width of $\neu_j$, respectively.
For the propagator we use the narrow width approximation. 	

Introducing a suitable set of spin vectors $s^a$ \cite{Haber94}, \cite{gudi99}, given in Appendix \ref{spinvectors},
the spin density matrices, eqs.~(\ref{rhop}) and (\ref{rhod}),
can be expanded in terms of the Pauli matrices $\tau^a,\ a=1,2,3$,
\begin{eqnarray}
	\rho^P_{\la_j \la_j^\prime} &=& 
		\delta_{\la_j \la_j^{\prime}}P + \sum_{a=1}^3 \tau_{\la_j \la_j^{\prime}}^a  {\Sigma}_P^a,
\label{rhoP}
\\
	\rho^D_{\la_j^\prime \la_j} &=& 
		\delta_{\la_j^\prime \la_j}D + \sum_{a=1}^3 \tau_{\la_j^{\prime} \la_j}^a  {\Sigma}_D^a.
\label{rhoD}
\end{eqnarray}
In eq.~(\ref{rhoP}) ${\Sigma}_P^3/P$ is the longitudinal polarization of the neutralino, 
 ${\Sigma}_P^1/P$ the transverse polarization in the production plane and
 ${\Sigma}_P^2/P$ that perpendicular to the production plane.
Inserting the density matrices, eqs.~(\ref{rhoP}) and (\ref{rhoD}), into eq.~(\ref{tsquare}) we obtain
\begin{eqnarray}
	|{{T}}|^2 &=&  2|\Delta(\neu_j)|^2 
		(P D + \sum_{a=1}^3 {\Sigma}_P^a {\Sigma}_D^a).
\label{eq:tsquare}
\end{eqnarray}
The first term in eq.~(\ref{eq:tsquare}) is independent of the neutralino polarization 
and the second term describes the spin correlation between production and decay of the neutralino $\neu_j$.

The resonant contributions from the s-channel exchange of $H$ and $A$ to $P$ and $\Sigma_P^a$ 
are denoted by $P_{{R}}$ and $\Sigma_{{R}}^a$, respectively.
Explicit expressions for $P_{{R}}$ and $\Sigma_{{R}}^3$
are given in the Appendix, 
eqs.~(\ref{pcpcons}) and (\ref{sigmacpcons}),
while the resonant contributions $\Sigma_{{R}}^1$ and $\Sigma_{{R}}^2$ to the transverse polarization of the neutralino vanish.
The non-resonant contributions to the production density matrix coefficients from $Z$ and slepton exchange
can be found in \cite{gudi99}.
The contributions from exchange of the lighter Higgs boson $h$ are numerically negligible. 
The interference of the chirality violating Higgs exchange amplitudes with chirality conserving
$Z$ and slepton exchange amplitudes is of order $m_\mu/\sqrt{s}$  
and can safely be neglected.
The coefficients $D$ and $\Sigma_D^3$ for the two-body decay of the neutralino into a lepton and a slepton 
are given in Appendix \ref{decdens}, eqs.~(\ref{DR}) and (\ref{SigmaD3}), for $\ell=e,\mu$ and eqs.~(\ref{Dtau}) and (\ref{SigmaD3tau}) for $\ell=\tau$.

The kinematical limits of the lepton energy in the CMS are \cite{Bartl:2003tr}
\begin{eqnarray}
	E_\ell^{max(min)}&=& 	 \bar{E}_\ell \pm \Delta_\ell ,
\end{eqnarray}with\begin{eqnarray}
	\bar{E}_\ell &=& \frac{E_\ell^{max}+E_\ell^{min}}{2} = \frac{ m_{\chi_j}^2-m_{\slepton}^2}{2 m_{\chi_j}^2} E_{\chi_j},
\label{ehalf}
\\
	\Delta_\ell &=& \frac{E_\ell^{max}-E_\ell^{min}}{2} = \frac{ m_{\chi_j}^2-m_{\slepton}^2}{2 m_{\chi_j}^2} |\vec{p}_{\chi_j}|.
\label{edif}
\end{eqnarray}
It is useful to introduce the average
\begin{eqnarray}
	\bar{P} 		&=& \frac{1}{4\pi}\int 	P 		d\Omega_{\neu},
\\
	\bar{\Sigma}^3_P	&=& \frac{1}{4\pi}\int {\Sigma}^3_P 	d\Omega_{\neu}, 
\label{eq:sigmabar}
\end{eqnarray}
over the neutralino production angles in the CMS. 
Then, the integrated cross section for neutralino production, eq.~(\ref{prod}), and subsequent leptonic decay
$\neu_j\to\ell^\pm\slepton^\mp_n$, with $n=R,L$ for $\ell=e,\mu$ ($n=1,2$ for $\ell=\tau$), 
is given by
\begin{eqnarray}
 \sigma_{\ell}^{n} &=& 
		\frac{1 }{64\pi^2}
                \frac{\sqrt{\lambda_{ij}}}{s^2}   \,
		\frac{(m_{\chi_j}^2-m_{\tilde{\ell}}^2)}{m_{\chi_j}^3 {\Gamma_{\chi_j}}}
\,      \bar{P} D.
\end{eqnarray}
Explicit expressions for $D$ are given in eqs.~(\ref{DR}) and (\ref{Dtau}) for $\ell=e,\mu$ and $\ell=\tau$, respectively,
and $\lambda_{ij}$ is the triangle function, defined in Section~\ref{section:couplings}.
The energy distribution of the lepton is 
\begin{equation}
\frac{d\sigma^{n}_{\ell^{\pm}}}{dE_\ell} =
	\frac{\sigma_\ell^{n}}
		{2\Delta_\ell}\left[ 1 + \,\etarl\eta_{\ell^\pm}\frac{\bar{\Sigma}_P^3}{\bar{P}} 
	\frac{(E_\ell - \bar{E_\ell})}{\Delta_\ell} 
					\right].
\label{edist2}
\end{equation}
Here $\eta_{\ell^\pm}= \mp1$.
Further, 
$\eta^R_{e,\mu}=+1$ and $\eta^L_{e,\mu}=-1$ for the decay into $\tilde{e}_R,\tilde{\mu}_R$ and $\tilde{e}_L,\tilde{\mu}_L$, respectively.
For the decay $\neu_j\to\tau^\pm\stau^\mp_{1,2}$ 
the lepton energy dependent term in (\ref{edist2}) is suppressed, due to stau mixing,
by
\begin{equation}
\etatauk= \frac{|b_{nj}^{\stau}|^2 - |a_{nj}^{\stau}|^2}
		{|b_{nj}^{\stau}|^2 + |a_{nj}^{\stau}|^2},
\label{eq:eta_rl}
\end{equation}
where the $\neu_j\stau_n\tau$ couplings $a_{nj}^{\stau}$ and $b_{nj}^{\stau}$ are defined in eq.~(\ref{eq:taucouplings}). 

Due to the Majorana character of the neutralinos,
the contribution to $\Sigma^3_P$ from the non-Higgs channels is forward-backward antisymmetric \cite{gudi.majorana},
whereas that from Higgs exchange is isotropic.
Then, the non-resonant contribution in eq.~(\ref{eq:sigmabar}) vanishes and, 
neglecting the interference of the resonant amplitudes with the $Z$ and slepton exchange amplitudes,
\begin{eqnarray}
	\bar{\Sigma}^3_P &=& {\Sigma}_{{R}}^3.
\label{sigmaPint}
\end{eqnarray}
From eq.~(\ref{sigmacpcons}) follows that ${\Sigma}_{{R}}^3$, and thus the energy dependent term in eq.~(\ref{edist2}), are proportional to the interference of the $H$ and $A$ exchange amplitudes.

\begin{figure}[t]
\centering
\begin{picture}(14,4.6)
\put(.7,-11.95){\includegraphics{fig.1}}
\put(5.93,3.){\color{blue3}$ \scriptstyle \ell^-$}
\put(5.02,3.){\color{green}$ \scriptstyle \ell^+$}
\put(8.1,0.85){\color{blue3}$ \scriptstyle \ell^+$}
\put(8.6,1.22){\color{green}$ \scriptstyle \ell^-$}
\put(1.4,3.85){$ \frac{1}{\sigma_\ell}\frac{d\sigma_\ell}{d E_\ell}{\scriptstyle[ GeV^{-1}]}$}
\put(8.38,-.15){$ \scriptstyle E_\ell[GeV]$}
\end{picture}
\caption{\small Normalized primary and secondary lepton energy distribution, 
with $\mathcal{A}_\ell^R=0.2$, 
$m_{\neu_2}=250$ GeV, $m_{\slepton_R}=200$ GeV, $m_{\neu_1}=60$ GeV and $\sqrt{s}=450$ GeV.
The dashed curves correspond to the decay chain $\neu_2\to\ell^-\slepton^+_R,\ \slepton^+_R\to\neu_1\ell^+$ 
and the dash-dotted curves to 
$\neu_2\to\ell^+\slepton^-_R,\ \slepton^-_R\to\neu_1\ell^-$ .
}
\label{fig:edist.l1l2}
\end{figure}

\subsection{Lepton energy distribution asymmetry}

For the processes $\mu^+\mu^-\to\neu_i\neu_j$ with subsequent decay 
$\neu_j\to\ell^+\slepton_{R,L}^-$, with $\ell=e,\mu$, and $\neu_j\to\tau^+\stau_{1,2}^-$, 
 as well as the charge conjugated decays,
we define the asymmetries $\mathcal{A}_{{\ell^+}}^{n}$ and $\mathcal{A}_{{\ell^-}}^{n}$, 
with $n=R,L$ for $\ell=e$ and $\ell=\mu$, and $n=1,2$ for $\ell=\tau$,
\begin{eqnarray}
	\mathcal{A}_{{\ell^\pm}}^{n} &=& 
	   \frac{\sigma_{\ell^\pm}^{n}(E_\ell>\bar{E}_\ell)-\sigma_{\ell^\pm}^{n}(E_\ell<\bar{E}_\ell)}
		{\sigma_{\ell^\pm}^{n}(E_\ell>\bar{E}_\ell)+\sigma_{\ell^\pm}^{n}(E_\ell<\bar{E}_\ell)}
\label{apoltot}
\\
&=& \frac{1}{2}\etarl\eta_{\ell^\pm}
	\frac{{\Sigma}_R^3}{\bar{P}}
\label{apoltot2}
\end{eqnarray}
in order to isolate the term dependent on the interference of $H$ and $A$ 
in eq.~(\ref{edist2}). 

The slepton 
decays subsequently into a neutralino and a secondary lepton. 
The latter needs to be identified from the primary lepton. 
Therefore, it is useful to define the asymmetry	
\begin{eqnarray}
	\mathcal{A}_{{\ell}}^{n} &=& \frac{1}{2}(\mathcal{A}_{{\ell^-}}^{n} - \mathcal{A}_{{\ell^+}}^{n}),
\label{apoltot+-}
\end{eqnarray}
equivalent to $\mathcal{A}_{{\ell^-}}^{n}$ since for the primary lepton $\mathcal{A}_{{\ell^-}}^{n} = - \mathcal{A}_{{\ell^+}}^{n}$.
The advantage of the new asymmetry, 	
eq.~(\ref{apoltot+-}),
is that the largest part of the non-irreducible background from the secondary lepton drops out
because its energy distribution is only weakly dependent on the sign of the lepton charge. 
In fig.~\ref{fig:edist.l1l2} we show the normalized energy distributions of the primary and secondary leptons for both charge cases, for a sample scenario with $\mathcal{A}_{{\ell}}^{R}=0.2$.

Denoting by $\sigma_{{R}}(\mu^+\mu^-\to\neu_i\neu_j)$ the resonant contribution to the 
production cross section $\sigma(\mu^+\mu^-\to\neu_i\neu_j)$  
we relate $\bar{P}$ to $P_{{R}}$ by
\begin{eqnarray}
	\bar{P} &=& 		
				\frac{\sigma(\mu^+\mu^-\to\neu_i\neu_j)}{\sigma_{{R}}(\mu^+\mu^-\to\neu_i\neu_j)} 
	P_{{R}},
\label{Pint}
\end{eqnarray}
and express $\mathcal{A}_\ell^n$ 
in the form
\begin{eqnarray}
	\mathcal{A}_{{\ell}}^{n} &=& 
\frac{1}{2}
	\,\etarl 
		\frac{\sigma_{{R}}(\mu^+\mu^-\to\neu_i\neu_j)}{\sigma(\mu^+\mu^-\to\neu_i\neu_j)} 
	\,{\mathcal{P}}_j^{{R}},
\label{apoltochipol}
\\	
	{\mathcal{P}}_j^{{R}}
	&=& 
	\frac{{\Sigma}_{{R}}^3}{{P_{{R}}}}.
\label{lambdaha}
\end{eqnarray}

The contribution of the $H$-$A$ interference to the asymmetry, eq.~(\ref{apoltochipol}), is
contained in the coefficient $\mathcal{P}_j^{{R}}$, which 
has the following dependence on the longitudinal beam polarizations $P^L_+$ of $\mu^+$ and $P^L_-$ of $\mu^-$:
\begin{eqnarray}
	{\mathcal{P}}_j^{{R}}
	&=&  \frac{P^L_+ +P^L_-}{1+P^L_+ P^L_-}
	{\mathcal{P}}_{j,R_+R_-}^{{R}},
\label{chipolmax}
\end{eqnarray}
where 
$ {\mathcal{P}}_{j,R_+R_-}^{{R}} =  {\mathcal{P}}_{j}^{{R}} $ 
for $P^L_+=P^L_-=1$, i.e.\ for right handed $\mu^+$ and $\mu^-$ beams.

Since ${\mathcal{P}}_j^{{R}}$ is proportional to the interference of the $H$ and $A$ exchange amplitudes 
a non-vanishing asymmetry of the lepton energy distribution is a clear indication of nearly degenerate scalar resonances with opposite $CP$ quantum numbers.

The statistical significance of the asymmetry $\mathcal{A}_\ell^n$	
can be defined by
\begin{eqnarray}
	\mathcal{S}_\ell^n
	&=& |\mathcal{A}^n_{\ell}| 
		\sqrt{2\sigma(\mu^+\mu^-\to\neu_i\neu_j)BR(\neu_j\to\ell^-\slepton_n^+){\mathcal{L}_{ef\!f}}},
\label{significance}
\end{eqnarray}
where $\mathcal{L}_{ef\!f} =\eps^n_\ell \mathcal{L}$ denotes the effective integrated luminosity, with $\eps^n_\ell$ the detection efficiency of the leptons 
in the processes $\neu_j\to\ell^\mp\slepton_n^\pm$ and 
$\mathcal{L}$ the integrated luminosity.

\section{Determination of the Higgs-neutralino couplings}
\label{section:couplings}
A measurement of the asymmetry of the primary lepton energy distribution 
opens the possibility to determine the 
$H$-neutralino and $A$-neutralino couplings using eqs.~(\ref{apoltochipol}) and (\ref{lambdaha}).
Inserting in eq.~(\ref{lambdaha}) the expressions of $P_R$ and $\Sigma_{R}^3$, eqs.~(\ref{pcpcons}) and (\ref{sigmacpcons}), 
we obtain,
for right handed $\mu^+$ and $\mu^-$
beams ($P_+^L=P_-^L=1$):
\begin{eqnarray}
{\mathcal{P}}_{j,R_+R_-}^{{R}}
 &=&
 \frac{
	2\gamma_{ij} 	\mbox{Re}(\Delta(H)\Delta^*(A)) \sqrt{s_{ij}^+s_{ij}^-}
	}
	{ r_{ij} \,|\Delta{(H)}|^2 \,s_{ij}^+ + r_{ij}^{-1} \, |\Delta{(A)}|^2\, s_{ij}^- } ,
\label{apol1}
\label{apol2}
\end{eqnarray}
where
\begin{eqnarray}
	\Delta(\phi) &=& i [(s-m_\phi^2) + i m_\phi \Gamma_\phi]^{-1},\qquad \phi=H,A,
\label{hpropagators}
\\
	s_{ij}^{\pm}&=& s -(\eta_i{m}_{\chi_i}\pm\eta_j {m}_{\chi_j})^2,
\label{s_pm}
\\
r_{ij}&=&\frac{|c_{ij}^{(H)}c^{(H\mu)}|}{|c_{ij}^{(A)} c^{(A\mu)}|},
\\
\gamma_{ij}&=&	\frac{\mbox{Im}(c_{ij}^{(H)}c_{ij}^{(A)\ast})}{|c_{ij}^{(H)}c_{ij}^{(A)}|} 
		\frac{\mbox{Im}(c^{(H\mu)}c^{(A\mu)\ast})}{|c^{(H\mu)}c^{(A\mu)} |}.
\end{eqnarray}
Since we assume $CP$ conservation $ \gamma_{ij}$ takes the values $\pm 1$ for interfering amplitudes with opposing $CP$ eigenvalues, 
as is here the case, and vanishes for interfering amplitudes of same $CP$.
Notice that the functions $s_{ij}^{\pm}$ depend on the relative $CP$ phase factor of the neutralinos $\eta_{ij}=\eta_i\eta_j$, 
with $\eta_i$, $\eta_j$ defined in Section~\ref{section:lagrdens}.

We obtain $\mathcal{P}_j^R$ from $\mathcal{A}_\ell^n$ using eq.~(\ref{apoltochipol}).
The neutralino-slepton couplings needed to evaluate $\eta_\ell^n$ will have been precisely studied at a linear collider, see, e.g., \cite{TDR,Nojiri.Boos.Martyn},
and the resonant cross section of neutralino production $\sigma_{{R}}(\mu^+\mu^-\to\neu_i\neu_j)$ 
can be obtained subtracting the continuum contributions from the integrated production cross section $\sigma(\mu^+\mu^-\to\neu_i\neu_j)$. 
The continuum can be estimated extrapolating the production cross sections measured below and above the resonance region \cite{Fraas:2003cx}.

It is possible to solve eq.~(\ref{apol2}) for 
$\gamma_{ij} r_{ij}$. 
To determine $\gamma_{ij}$ and $r_{ij}$ the resonance parameters, as well as the peak and off resonance cross sections, need to be precisely known.
The widths and masses of nearly degenerate Higgs resonances with different CP quantum numbers may be determined 
with transverse beam polarization, 
which enhances or suppresses the Higgs boson cross section depending on the Higgs $CP$ quantum numbers \cite{poltransv}. 

The product of couplings
\begin{equation}
	k_{ij}=\mbox{Im}(c_{ij}^{(H)}c_{ij}^{(A)\ast}) \mbox{Im}(c^{(H\mu)}c^{(A\mu)\ast})
\end{equation}
can be determined with a measurement of $\mathcal{A}_\ell^n\,\sigma(\mu^+\mu^-\to \neu_i\neu_j)$:
\begin{eqnarray}
	\mathcal{A}_{{\ell}}^{n}\, {\sigma(\mu^+\mu^-\to\neu_i\neu_j)} &=& 
	\frac{1}{2}\,\etarl {\sigma_{{R}}(\mu^+\mu^-\to\neu_i\neu_j)}\frac{{\Sigma}_{{R}}^3}{{P_{{R}}}}
\nonumber\\
&=&	
	(2-\delta_{ij})
	\frac{g^4}{16\pi}
	\frac{{\lambda_{ij}}}{s} 
	(P_+^L + P_-^L) 
		\mbox{Re} \{ (\Delta(H))(\Delta(A))^*\} 
	\eta_j \etarl k_{ij}, \qquad
\end{eqnarray}
where in the last equality we used the relation
\begin{eqnarray}
{\sigma}_{{R}}(\mu^+\mu^-\to\neu_i\neu_j) 
&=& \frac{\sqrt{\lambda_{ij}}}{8\pi s^2} P_R, 
\end{eqnarray}
the triangle function is defined by $\lambda_{ij}={s_{ij}^+s_{ij}^-}$, and $P_R$ and $\Sigma_{R}^3$ are given in eqs.~(\ref{pcpcons}) and (\ref{sigmacpcons}), respectively.

Notice that the $H$-$A$ interference in neutralino production depends on $\gamma_{ij}$,
while pure $H$ or $A$ exchange does not. 
Therefore, a measurement of the lepton energy asymmetries provides unique information on the Higgs-neutralino couplings.


\section{Numerical results}

We present numerical results for the neutralino production cross sections $\sigma(\mu^+\mu^-\to\neu_i\neu_j)$,
the asymmetries $\mathcal{A}^{R}_{\ell}$ and $\mathcal{A}^1_\tau$ of the lepton energy distribution
and the statistical significance at center of mass energies
around the resonances of the neutral Higgs bosons $H$ and $A$.
We study the dependence on the MSSM parameters $\tan\beta$, $\mu$, $M_2$ and $m_A$
in the mixed scenarios {\bf B} and in the gaugino-like scenario {\bf SPS1a} defined in Table~\ref{scenarios1}. 
Further, we discuss in scenario {\bf SPS1a} \cite{Allanach:2002nj} the influence of beam polarization.

\begin{table}
\renewcommand{\arraystretch}{1.2}
\begin{center}
\begin{tabular}{|c||cccc||cccc|}
\hline
{\bf Scenarios} & {\bf SPS1a}& {\ \bf B5\ }& {\ \bf B10\ }& {\ \bf B20\ }& {\ \bf B5'\ \ }& {\ \bf B10'\ }  & {\ \bf B5''\ }& {\ \bf B10''\  }
\\
\hline
\hline
$\tan\beta$ 			&10	&5 	&10	&20	&5 	&10	&5 	&10	\\
\hline
$M_2$[GeV] 			&192.7	&280	&280	&280	&280	&280	&280	&280	\\
\hline
$\mu$[GeV] 			&352.4	&250	&250	&250	&250	&250	&250	&250	\\
\hline
$m_{\neu_3}$[GeV]		&359	&255 	&257	&258	&255 	&257	&255 	&257	\\
\hline
$m_{\neu_2}$[GeV]		&177	&209 	&212	&214	&209 	&212	&209 	&212	\\
\hline
$m_{\neu_1}$[GeV] 		&96	&128	&131	&132	&128	&131	&128	&131	\\
\hline
\hline
$m_0$[GeV]			&100 	&100	&100	&100	&100	&100	&100	&100	\\
\hline
$m_{\tilde{e}_R}$[GeV] 		&143	&173 	&173	&173	&173 	&173	&173 	&173	\\
\hline
\hline
$m_A$[GeV] 		&393.6	&450 	&450	&450	&350 	&350	&550	&550	
\\
\hline
$m_H$[GeV] 		&394.1	&451.4	&450.4	&450.1	&351.9	&350.5	&551.1	&550.3
\\
\hline
$\Gamma_A$[GeV]		&1.38	&2.29	&2.11	&4.33	&0.43	&0.82	&3.63	&3.34
\\
\hline
$\Gamma_H$[GeV]		&0.93	&1.12	&1.33	&3.68	&0.27	&0.71	&2.83	&2.76
\\
\hline
\end{tabular}
\end{center}
\caption{Reference scenarios. 
GUT relations are assumed for the gaugino soft breaking mass parameters $M_1=5/3 \tan^2\theta_W M_2$ 
and for the slepton mass parameters \cite{Hall:zn}.
The resonance parameters are evaluated with HDECAY \cite{HDECAY}.}
\renewcommand{\arraystretch}{1.0}
\label{scenarios1}
\end{table}

In order to reduce the number of parameters 
we assume the GUT relations for the gaugino mass parameters, 
$M_1=5/3\tan^2\theta_W M_2$, 
and for the slepton masses of the first two generations, $\ell=e,\mu$, \cite{Hall:zn}
\begin{eqnarray}
	m_{\slepton_R}^2&=& m_0^2+m_\ell^2+0.23 M_2^2 - m_Z^2 \cos 2\beta \sin^2\theta_W,
\label{mslr}
\\
	m_{\slepton_L}^2&=& m_0^2+m_\ell^2+0.79 M_2^2 - m_Z^2 \cos 2\beta (\frac{1}{2}-\sin^2\theta_W).
\label{msll}
\end{eqnarray}
where $m_Z$ is the mass of the $Z$ boson and $m_0$ is the scalar mass parameter. 
The masses of the stau mass eigenstates $\stau_1$ and $\stau_2$ are obtained 
diagonalizing the stau mass matrix, eq.~(\ref{eq:mm}). 
Our scenarios have been chosen such that
the mass of the lightest stau, $\stau_1$, is of order $m_{\slepton_R}$, and the mass of the heavier stau, $\stau_2$, is of order $m_{\slepton_L}$,
and $m_{\slepton_R}<m_{\chi_j}<m_{\slepton_L}$.
Therefore, 
neglecting the three-body decays of $\neu_j$, 
only $\neu_j\to\ell\slepton_R,\ \ell=e,\mu$ and $\neu_j\to\tau\stau_1$ contribute to the energy spectrum of the leptons. 
We show
the neutralino branching ratios into lepton slepton pairs in Table~\ref{neu.br2}. 
Details of stau mixing can be found in Appendix~\ref{tau.neutralino}. 

\begin{table}
\renewcommand{\arraystretch}{1.2}
\begin{center}
\begin{tabular}{|c||cccc|}
\hline
{\bf Scenarios} & {\bf SPS1a} & {\ \ \bf B5\ } & {\ \bf B10\ } & {\ \bf B20\ } 
\\
\hline
\hline
$A_{\tau}$[GeV] 				& -254	& 0	& 0	& 0	\\
\hline
BR$(\neu_2\to\ell^-\slepton_R^+)[\%]$ 		&3.2	& 16.3	& 15.2	& 11.3	\\
\hline
BR$(\neu_2\to\tau^-\stau_1^+)[\%]$ 		&42.5	& 17.3 	& 19.6	& 27.4	\\
\hline
\end{tabular}
\end{center}
\renewcommand{\arraystretch}{1.0}
\caption{Neutralino branching ratios, $\ell=e,\mu$.}
\label{neu.br2}
\end{table}


\subsection{$\neu_1\neu_2$ production}
%
We first discuss, for $\neu_1\neu_2$ production, 
the dependence of the asymmetries and cross sections on the beam polarization
in scenario {\bf SPS1a}, the dependence on $\tan\beta$ 
in scenarios {\bf B5}, {\bf B10} and {\bf B20} 
and the dependence on $m_A$ in scenarios {\bf B5'},{\bf B5''}, {\bf B10'} and {\bf B10''}. 
All the scenarios are defined in Table~\ref{scenarios1}

\subsubsection{Beam polarization}
\begin{figure}[ht]
\centering
\begin{picture}(14,4.7)
\put(-2,-9.3){\includegraphics{fig.2a}}
\put(3.2,-9.3){\includegraphics{fig.2b}}
\put(8.1,-9.3){\includegraphics{fig.2c}}
\put(.92,4.04){\small (a)}
\put(6.12,4.04){\small (b)}
\put(11.02,4.04){\small (c)}
\put(-.3,4.1){$ \scriptstyle \mathcal{A}_{\ell}^R$}
\put(4.7,4.1){$ \scriptstyle \sigma[fb]$}
\put(2.7,-.1){$ \scriptstyle \sqrt{s}[GeV]$}
\put(7.9,-.1){$ \scriptstyle \sqrt{s}[GeV]$}
\put(10.05,4.1){$  \scriptstyle \mathcal{S}_\ell^R$}
\put(12.8,-.1){$ \scriptstyle \sqrt{s}[GeV]$}
\end{picture}
\caption{\small 
$\mu^+\mu^- \to \neu_1 \neu_2$, $\neu_2\to\ell^-\slepton_R^+$, $\ell=e,\mu$, for scenario {\bf SPS1a}. 
(a): $\mathcal{A}^R_{\ell}$, (b): neutralino production cross section and (c): significance with luminosity times detection efficiency $\eps \mathcal{L}=\mathcal{L}_{ef\!f}=0.5 fb^{-1}$ (for $\neu_2\to\ell^-\slepton_R^+$), for
beam polarizations: $P^L_+=P^L_-=-0.2$ (dash-dotted), -0.3 (dashed), and -0.4 (solid).
        }
\label{SPS1a}
\end{figure}
%
In figs.~\ref{SPS1a}a and \ref{SPS1a}b we show 
	the asymmetry $\mathcal{A}^R_{\ell}$, $\ell=e,\mu$,
and	the cross section $\sigma(\mu^+\mu^-\to\neu_1 \neu_2)$, respectively,
for scenario {\bf SPS1a} as a function of the center of mass energy around the heavy Higgs resonances.
Since the resonances are completely overlapping 
the interference between the $CP$ even and $CP$ odd amplitudes is large,
resulting in large asymmetries in the resonance region.
The largest asymmetries are found at $\sqrt{s}\simeq m_H$
where the $CP$ even and $CP$ odd amplitudes are of the same order,
because,
due to the relative $CP$ phase factor $\eta_{12}\equiv\eta_1\eta_2=1$ of the neutralinos, 
the $CP$ even amplitudes are P-wave suppressed. 
The largest production cross sections are found at $\sqrt{s}\simeq m_A$. 

We show the cross section, the asymmetry and the significance of scenario {\bf SPS1a} 
for $P_+^L=P_-^L=-0.2,\ -0.3$, $-0.4$.
The dependence on the longitudinal beam polarization of the resonant cross section is given by the factor $1+P^L_+ P^L_-$, and is rather 
weak for the polarization degrees expected at a muon collider.
From eqs.~(\ref{apoltochipol}) and (\ref{chipolmax}) follows that the polarization dependence of the asymmetry is then roughly $\mathcal{A}_\ell^n\sim P^L_+ + P^L_-$. 
For the statistical significance, defined in eq.~(\ref{significance}), follows then $\mathcal{S}_\ell^n\sim P^L_+ + P^L_-$.
\subsubsection{Stau mixing dependence}
\label{sec:staudep}
The asymmetry for the $\tau$ energy spectrum depends strongly on the mixing in the stau sector.
The $\tau$ energy asymmetry is obtained by $\mathcal{A}_\tau^1=\etatau \mathcal{A}_\ell^R$, eqs.~(\ref{apoltochipol}) and (\ref{eq:eta_rl}). 
For the {\bf SPS1a} scenario $\etatau=-0.87$.
Notice that the asymmetries $\mathcal{A}_\tau^1$ and $\mathcal{A}_\ell^R$ have opposite signs.
The marked difference between $\mathcal{A}_\tau^1$ and $\mathcal{A}_\ell^R$ is 
due to stau mixing, which allows the lightest scalar tau $\stau_1$ to have a large left component.
For the gaugino-like {\bf SPS1a} scenario the second lightest neutralino $\neu_2$ is wino-like, and thus
has large left handed couplings to lepton-slepton pairs.
Therefore, $\neu_2$ decays dominantly into $\tau$-$\stau_1$ pairs, 
see the branching ratios for {\bf SPS1a} in Table~\ref{neu.br2}.
For $A_\tau=\mu\tan\beta$
the stau mass matrix is diagonal, eq.~(\ref{eq:mm}). 
Then, the branching ratios for the decays $\neu_2\to\ell\slepton_R$ and $\neu_2\to\tau\stau_1$
are comparable in size. 
However, $\mathcal{A}_\tau^1$ is still smaller than $\mathcal{A}_\ell^R$, 
with $\etatau=0.53$, 
due of the larger couplings to the higgsino components to the scalar taus.

In the resonance region we find, for $P^L_+=P^L_-=-0.3$,  
$\mathcal{S}_\ell^R\simeq 1.5 \sqrt{\mathcal{L}_{ef\!f} [fb^{-1}]}$, $\ell=e,\mu$, 
and $\mathcal{S}_\tau^1\simeq 4.5 \sqrt{\mathcal{L}_{ef\!f}[fb^{-1}]}$.
In fig.~\ref{SPS1a}c we show the statistical significance, defined in eq.~(\ref{significance}), for $\mathcal{A}^R_{\ell}, \ \ell=e,\mu$ 
with an effective integrated luminosity $\mathcal{L}_{ef\!f}=0.5 fb^{-1}$.

\subsubsection{$\tan\beta$ dependence} 
\label{sec:tbdep}
\begin{figure}[ht]
\centering
\begin{picture}(14,4.7)
\put(-2,-9.3){\includegraphics{fig.3a}}
\put(3.2,-9.3){\includegraphics{fig.3b}}
\put(8.1,-9.3){\includegraphics{fig.3c}}
\put(10.0,4.1){$  \scriptstyle \mathcal{S}_\ell^R$}
\put(12.9,-.1){$ \scriptstyle \sqrt{s}[GeV]$}
\put(.92,4.03){\small (a)}
\put(6.12,4.03){\small (b)}
\put(11.02,4.03){\small (c)}
\put(-.4,4.1){$ \scriptstyle \mathcal{A}_{\ell}^R$}
\put(4.6,4.1){$ \scriptstyle \sigma[fb]$}
\put(2.7,-.1){$ \scriptstyle \sqrt{s}[GeV]$}
\put(7.9,-.1){$ \scriptstyle \sqrt{s}[GeV]$}
\end{picture}
\caption{\small 
$\mu^+\mu^- \to \neu_1 \neu_2$, $\neu_2\to\ell^-\slepton_R^+$, $\ell=e,\mu$ for scenarios {\bf B5}, {\bf B10} and {\bf B20}. 
(a): $\mathcal{A}^R_{\ell}$, (b): neutralino production cross section and (c): statistical significance with luminosity times detection efficiency $\eps \mathcal{L}=\mathcal{L}_{ef\!f}=0.5 fb^{-1}$ and 
$P^L_+=P^L_-=-0.3$. 
$\tan\beta=5$ (dash-dotted), 10 (dashed),
and 20 (solid).
        }
\label{B.tb}
\end{figure}   
In fig.~\ref{B.tb}a 
we show the asymmetry $\mathcal{A}^R_{\ell},\ \ell=e,\mu$, 
for scenarios {\bf B5, B10} and {\bf B20}, 
for $P_+^L=P_-^L=-0.3$.
These scenarios differ only by the value of $\tan\beta$.
For increasing $\tan\beta$
the mass difference $m_H-m_A$ decreases and the widths $\Gamma_H$ and $\Gamma_A$ increase.
This results in a larger overlap of the resonances which leads to
large asymmetries in the resonant region. 
For $\tan\beta=5$, with only partial overlap of the resonances, 
the asymmetry is further suppressed by the relative larger continuum contribution to the cross section due to the smaller Higgs-muon couplings.
However, it shows an interesting energy dependence due to the different complex phases of the Breit-Wigner propagators of $H$ and $A$.
The maximum of the asymmetry is found at $\sqrt{s}\simeq m_H$, as already discussed for scenario {\bf SPS1a}.

In fig.~\ref{B.tb}b we show the cross sections $\sigma(\mu^+\mu^-\to\neu_1\neu_2)$ for scenarios {\bf B5, B10} and {\bf B20}.
The largest peak cross sections are found for $\tan\beta=10$. 
For $\tan\beta=5$ the resonant cross sections are suppressed by the smaller Higgs-muon couplings, 
while for $\tan\beta=20$ 
they are suppressed by the larger resonance widths.

In the resonant region we find, for $\ell=e,\mu$,
$\mathcal{S}_\ell^R\ge 3\sqrt{\mathcal{L}_{ef\!f}[fb^{-1}]}$ for $\tan\beta=10$ and $\tan\beta=20$, 
while for $\tan\beta=5$ the statistical significances reach $\mathcal{S}_\ell^R\simeq \sqrt{\mathcal{L}_{ef\!f}[fb^{-1}]}$ at $\sqrt{s}\simeq m_H$. 
In fig.~\ref{B.tb}c we show the statistical significances for an effective integrated luminosity $\mathcal{L}_{ef\!f}=0.5 fb^{-1}$.

The effect of stau mixing on the asymmetry $ \mathcal{A}_{\tau}^1 = \eta^1_\tau \mathcal{A}_{\ell}^R$,
with $ \eta^1_\tau$ defined in eq.~(\ref{eq:eta_rl}),
increases with $\tan\beta$.
It is weaker in the mixed scenarios than in the gaugino-like {\bf SPS1a} scenario, 
as can also be observed comparing the neutralino branching ratios of Table~\ref{neu.br2}.
We find,
for scenarios {\bf B5, B10} and {\bf B20}, 
$\etatau=0.88$, 0.49 and -0.31, 
respectively.
The statistical significance for $\mathcal{A}_\tau^1$ is obtained from eq.~(\ref{significance}), where
the branching ratios of $\neu_2\to\ell^\mp\slepton^\pm_R,\ \ell=e,\mu$, and $\neu_2\to\ell^\mp\stau^\pm_1$ are shown in Table~\ref{neu.br2}.

For $A_{\tau}=\mu\tan\beta$, i.e.\ for a diagonal stau mass matrix with $\stau_1=\stau_R$,
we find $\etatau=0.96$, 0.79 and 0.30 for scenarios {\bf B5, B10} and {\bf B20}, respectively.

\subsubsection{$m_A$ dependence} 
\begin{figure}[ht]
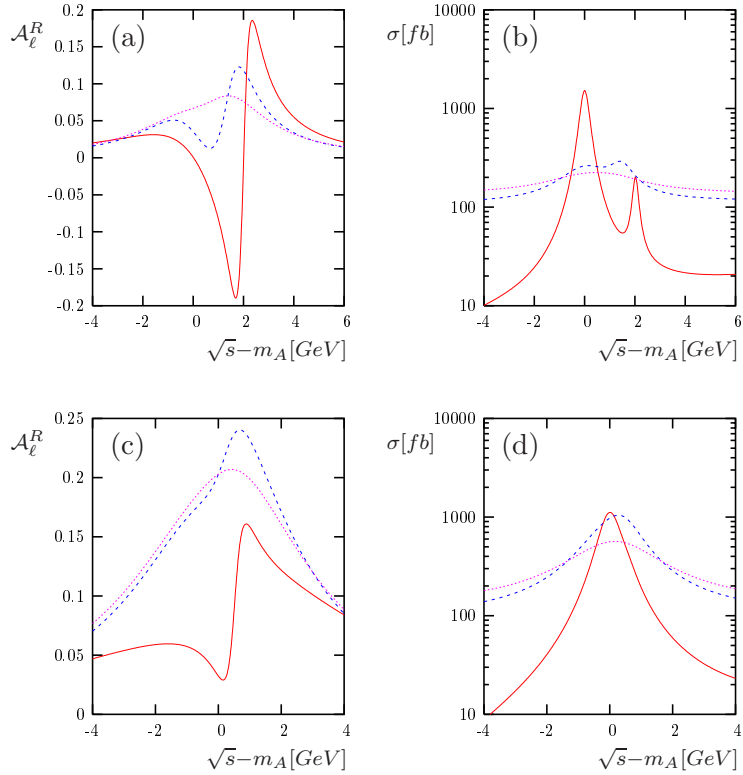

\centering
\begin{picture}(9,4.7)
\put(-2,-9.3){\includegraphics{fig.4a}}
\put(3.2,-9.3){\includegraphics{fig.4b}}
\put(.92,4.03){\small (a)}
\put(6.12,4.03){\small (b)}
\put(-.4,4.1){$ \scriptstyle \mathcal{A}_{\ell}^R$}
\put(4.6,4.1){$ \scriptstyle \sigma[fb]$}
\put(2.2,-.1){$ \scriptstyle \sqrt{s}-m_A[GeV]$}
\put(7.4,-.1){$ \scriptstyle \sqrt{s}-m_A[GeV]$}
\end{picture}
\begin{picture}(9,5.4)
\put(-2,-9.3){\includegraphics{fig.4c}}
\put(3.2,-9.3){\includegraphics{fig.4d}}
\put(.92,4.03){\small (c)}
\put(6.12,4.03){\small (d)}
\put(-.4,4.1){$ \scriptstyle \mathcal{A}_{\ell}^R$}
\put(4.6,4.1){$ \scriptstyle \sigma[fb]$}
\put(2.2,-.1){$ \scriptstyle \sqrt{s}-m_A[GeV]$}
\put(7.4,-.1){$ \scriptstyle \sqrt{s}-m_A[GeV]$}
\end{picture}
\caption{\small 
$\mu^+\mu^- \to \neu_1 \neu_2$,  $\neu_2\to\ell^-\slepton_R^+$, $\ell=e,\mu$, with 
	$P^L_+=P^L_-=-0.3$,
for scenarios {\bf B5, B10} 
with $m_A=450$ GeV (dashed),
{\bf B5', B10'} with $m_A=350$ GeV (solid) and
{\bf B5'', B10''} with $m_A=550$ GeV (dotted).
(a): $\mathcal{A}^R_{\ell}$ for $\tan\beta=5$, (b): neutralino production cross section for $\tan\beta=5$,
(c): $\mathcal{A}^R_{\ell}$ for $\tan\beta=10$, (d): neutralino production cross section for $\tan\beta=10$.
        }
\label{B.ma}
\end{figure}

In fig.~\ref{B.ma}a 
we compare the asymmetries $\mathcal{A}^R_{\ell},\ \ell=e,\mu$, 
for scenarios {\bf B5', B5} and {\bf B5''}, 
with different values of $m_A$,
as a function of $\sqrt{s}-m_A$, for 
$P_+^L=P_-^L=-0.3$.
In fig.~\ref{B.ma}b we show the corresponding cross section $\sigma(\mu^+\mu^-\to\neu_1\neu_2)$.
For larger Higgs masses their widths increase, 
and thus the interference
of the $H$ and $A$ exchange amplitudes.
However, the asymmetries are reduced by 
the larger continuum contribution to the cross section.

For smaller Higgs masses, here for $m_A=350$ GeV, threshold effects are stronger.
Since 
$\eta_{12}=1$,
the asymmetries nearly vanish for $\sqrt{s}\simeq m_A$, where the largest cross sections are found,
while the largest asymmetries are found at $\sqrt{s}\approx m_H$.
The asymmetries change sign between the two resonances, due to the complex phases of the propagators, 
and the maxima of $|\mathcal{A}_\ell^R|$ are found at center of mass energies slightly above and below $m_H$
and not on top of the $CP$ even resonance.
For larger values of $m_A$, here for $m_A=550$ GeV, 
the peak cross sections are suppressed by the larger widths.
To a lesser degree, they are enhanced 
by the larger phase space for neutralino production.

In figs.~\ref{B.ma}b and \ref{B.ma}d we show the analogous figures for scenarios {\bf B10', B10} and {\bf B10''}, 
with $\tan\beta=10$.
The effect of larger Higgs masses is weaker than for $\tan\beta=5$ because the overlap of the resonances is already large for $m_A=450$ GeV. 

The statistical significances at the center of mass energies where $|\mathcal{A}_\ell^R|$ is maximal 
is $\mathcal{S}_\ell^R\sim 0.8\sqrt{\mathcal{L}_{ef\!f}[fb^{-1}]}$ for scenario {\bf B5'} and
$\mathcal{S}_\ell^R\sim 1.4\sqrt{\mathcal{L}_{ef\!f}[fb^{-1}]}$ at $\sqrt{s}\simeq m_H$ for scenario {\bf B10'}.

\subsection{$\neu_1\neu_2, \neu_2\neu_2$ and $\neu_1\neu_3$ production}
\begin{figure}[h]
\centering
\begin{picture}(9,4.7)
\put(-2,-9.3){\includegraphics{fig.5a}}
\put(3.2,-9.3){\includegraphics{fig.5b}}
\put(.92,4.03){\small (a)}
\put(6.12,4.03){\small (b)}
\put(1.5,3.0){\color{red1}\tiny $\neu_1\neu_2$}
\put(.8,1.66){\color{green2}\tiny $\neu_2\neu_2$}
\put(1.8,1.1){\color{blue00}\tiny $\neu_1\neu_3$}
\put(-.4,4.1){$ \scriptstyle \mathcal{A}_{\ell}^R$}
\put(4.6,4.1){$ \scriptstyle \sigma[fb]$}
\put(2.7,-.1){$ \scriptstyle \sqrt{s}[GeV]$}
\put(7.9,-.1){$ \scriptstyle \sqrt{s}[GeV]$}
\end{picture}
\caption{\small 
$\mu^+\mu^- \to \neu_i \neu_j$, $\neu_j\to\ell^-\slepton_R^+$, $\ell=e,\mu$, for scenario {\bf B10}
with $i=1,\ j=2$ (solid), $i=1,\ j=3$ (dash-dotted) and $i=2,\ j=2$ (dashed). 
(a): $\mathcal{A}^R_{\ell}$ and (b): neutralino production cross section,
with $P^L_+=P^L_-=-0.3$. 
}
\label{B.ij}
\end{figure}
In fig.~\ref{B.ij} we show, for scenario {\bf B10}, the asymmetry $\mathcal{A}^R_{\ell}, \ell=e,\mu,$ 
and the cross sections $\sigma(\mu^+\mu^-\to\neu_i\neu_j)$ 
for the kinematically allowed pairs $\neu_1\neu_2,\ \neu_1\neu_3$ and $\neu_2\neu_2$, for $P_+^L=P_-^L=-0.3$.
Threshold effects are stronger in $\neu_2\neu_2$ and $\neu_1\neu_3$ production 
than in $\neu_1\neu_2$ production.
Therefore the P-wave suppression of
the $CP$ even ($CP$ odd) amplitude 
for $\neu_2\neu_2$, $( \neu_1\neu_3)$ production is stronger,
since $\eta_{22}=1$ $(\eta_{13}=-1)$.
The asymmetries for $\neu_1\neu_2$ and $\neu_2\neu_2$ production, however, 
are comparable in size, 
since the continuum contribution to $\neu_2\neu_2$ production is very small.
For $\neu_1\neu_3$ production, the asymmetry is smaller due to the interplay of the widths
and the amplitudes, with $\Gamma_H<\Gamma_A$ and
$A$ exchange suppressed, which results in a smaller interference of the two amplitudes.  
Notice also the different energy dependence of the asymmetry, 
with maxima at $\sqrt{s}< m_A$ and $\sqrt{s}> m_H$,
and of the cross section, with a maximum at $\sqrt{s}\simeq m_H$.

For the asymmetry in $\neu_2\neu_2$ production 
the statistical significance for $\ell=e,\mu$ in the resonance region is $\mathcal{S}_\ell^R\simeq 2\sqrt{\mathcal{L}_{ef\!f} [fb^{-1}]}$.
For $\neu_1\neu_3$, 
the statistical significance for $\ell=e,\mu$ is significantly smaller, of order $ \mathcal{S}_\ell^R\simeq 0.6\sqrt{\mathcal{L}_{ef\!f} [fb^{-1}]}$ in the resonance region,
because 
the branching ratios of $\neu_3$ into lepton and slepton pairs are strongly suppressed by the competing decay channels $\neu_3\to Z\neu_1$ and $\neu_3\to h \neu_1$,
with $BR(\neu_3\to\ell\slepton_R)=1\%$ for $\ell=e,\mu,$ and $BR(\neu_3\to\tau\stau_1)\simeq 5\%$. 

\subsection{$\mu-M_2$ plane}

\begin{figure}[ht]
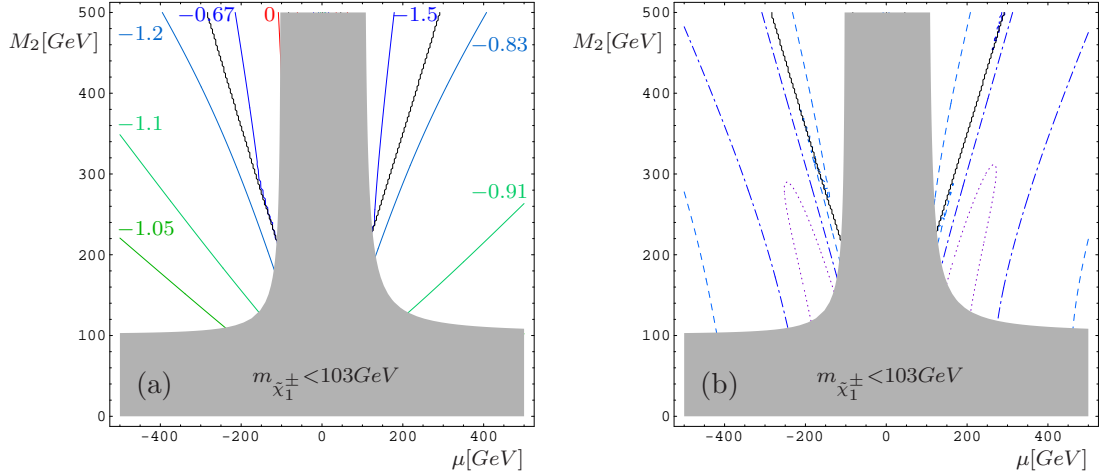

\centering
\begin{picture}(15.,7.)
\put(.5,0.2){\includegraphics{fig.6a}}
\put(8.,0.2){\includegraphics{fig.6b}}
\put(-.4,5.6){$  \scriptstyle M_2 [GeV]$}
\put(5.5,.05){$ \scriptstyle \mu [GeV]$}
\put(7.1,5.6){$  \scriptstyle M_2 [GeV]$}
\put(13.,.05){$ \scriptstyle \mu [GeV]$}
\put(1.3,1){\small (a)}
\put(8.8,1){\small (b)}
\put(1.85,5.95){\color{blue}$ \scriptstyle -0.67$}
\put(3.,5.95){\color{red}$ \scriptstyle 0$}
\put(4.7,5.95){\color{blue}$ \scriptstyle  -1.5$}
\put(1.05,5.7){\color{blue3}$ \scriptstyle -1.2$}
\put(5.75,5.55){\color{blue3}$ \scriptstyle -0.83$}
\put(5.75,3.6){\color{green4}$ \scriptstyle -0.91$}
\put(1.05,3.1){\color{green3}$ \scriptstyle -1.05$}
\put(1.05,4.5){\color{green4}$ \scriptstyle -1.1$}
\put(2.85,1.15){$ \scriptstyle m_{\tilde{\chi}_1^\pm}<103 GeV$ }
\put(10.35,1.15){$ \scriptstyle m_{\tilde{\chi}_1^\pm}<103 GeV$ }
\end{picture}
\caption{
	Contours of constant $\gamma_{12} r_{12}$ (a)
and $\mbox{Im}(c_{ij}^{(H)}c_{ij}^{(A)\ast}) $ (b),
for $\tan\beta=10$ and $m_A=450$ GeV.
In (b) we show
$\mbox{Im}(c_{ij}^{(H)}c_{ij}^{(A)\ast})=0.01$ (dashed), $0.03$ (dash-dotted) and $0.05$ (dotted). 
The wiggly lines in both figures indicate the level crossing of the $\neu_2$ and $\neu_3$ states, with
 $\eta_{12}=1$ in the area below and $\eta_{12}=-1$ in the area above the level crossing line.
The gray area is 
experimentally excluded by $m_{\tilde{\chi}_1^\pm} < 103$ GeV.
        }
\label{rkcontours}
\end{figure}

The MSSM parameters $\mu$ and $M_2$ affect strongly the neutralino couplings both to the Higgs bosons as to the lepton-slepton pairs.
The lepton energy asymmetries for neutralino decays into leptons of the first two families, $e$ and $\mu$, do not
depend on the couplings, while the dependence 
of stau mixing on the neutralino character has been briefly discussed in Section~\ref{sec:tbdep}.
The neutralino couplings to $H$ and $A$ are both enhanced in mixed scenarios since Higgs bosons couple to a higgsino-gaugino pair.
Therefore, the Higgs boson widths, and thus the interference of the resonances, are also enhanced.
In figs~\ref{rkcontours}a and \ref{rkcontours}b we show
contours in the $\mu-M_2$ plane of constant $\gamma_{12} r_{12}$ and $\mbox{Im}(c_{ij}^{(H)}c_{ij}^{(A)\ast}) $, respectively,
for $\tan\beta=10$ and $m_A=450$ GeV.
Notice
that $\gamma_{12}$ is negative in most of the experimentally allowed parameter space. 
Therefore the sign of the asymmetries for the first two lepton families constitutes a test of the Higgs-neutralino couplings in the MSSM.

The same qualitative dependence of $\gamma_{12} r_{12}$ and $k_{12}$ on $\mu$ and $M_2$
is found for different values of $\tan\beta$ and $m_A$.

\section{Summary and conclusion}
We have discussed the interference of the $CP$ even and the $CP$ odd amplitudes of 
                     neutral Higgs boson s-channel exchange in 
	$\mu^+\mu^-\to\neu_i\neu_j$
	with longitudinally polarized beams
in the $CP$ conserving MSSM.
To study this interference we
use 
the energy distribution of the 
lepton from the decay 
$\neu_j\to\ell^{\pm}\slepton^{\mp}$, $\ell=e,\mu,\tau$.
The asymmetry of the lepton energy distribution depends on the longitudinal polarization of the neutralinos, averaged over their production angles.
Since the neutralino longitudinal polarization is correlated to the longitudinal beam polarization when the $H$ and $A$ exchange amplitudes interfere, and vanishes otherwise,
the asymmetries can be used 
to determine the couplings of the $H$ and $A$ bosons to the produced neutralinos times, respectively,
their couplings to the muons. 
In particular, the sign of the asymmetry is sensitive to
the sign of the product of couplings of the neutralinos and of the muons to the Higgs bosons.

For a set of scenarios
we have analyzed the lepton energy asymmetries for $\neu_i\neu_j$ production with subsequent two-body decay, 
with emphasis on $\neu_1\neu_2$ production.
We find large asymmetries for 
nearly degenerate heavy neutral Higgs bosons and intermediate values of $\tan\beta$ and $m_A$.
Especially for $\neu_1\neu_2$ production 
we find statistical significances which would allow to measure the asymmetries at a muon collider. 

\section{Acknowledgments} 
We thank O.~Kittel for valuable comments and discussions.
This work was supported by the 'Deutsche Forschungsgemeineschaft' (DFG) under contract FR 1064/5-2.

\vspace{.8cm}

\begin{appendix}
\noindent{\Large\bf Appendix}
\setcounter{equation}{0}
\renewcommand{\thesubsection}{\Alph{section}.\arabic{subsection}}
\renewcommand{\theequation}{\Alph{section}.\arabic{equation}}
 
\section{Stau-neutralino couplings}
\setcounter{equation}{0}
\label{tau.neutralino}
The stau-neutralino couplings, defined by the interaction Lagrangian 
\begin{eqnarray}
	& & {\mathcal L}_{\tau\tilde{\tau} \chi_j }=  g\tilde \tau_n \bar \tau
		(a^{\tilde \tau}_{nj} P_R+b^{\tilde \tau}_{nj} P_L)\chi^0_j + {\rm h.c.}~,
 \quad n = 1,2; \; j=1,\dots,4, 
\end{eqnarray}
are \cite{Bartl:2002bh}
\begin{equation}
\label{eq:taucouplings}
	a_{nj}^{\tilde \tau}=
	({\mathcal R}^{\tilde \tau}_{nm})^{\ast}{\mathcal A}^\tau_{jm},\qquad
	b_{nj}^{\tilde \tau}=
	({\mathcal R}^{\tilde \tau}_{nm})^{\ast}{\mathcal B}^\tau_{jm},\qquad m=L,R
\end{equation}
with ${\mathcal R}^{\tilde \tau}_{nm}$ the stau mixing matrix defined in eq.~(\ref{eq:rtau})
and
\begin{equation}
	{\mathcal A}^{\tau}_j=\left(\begin{array}{ccc}
	f^{L}_{\tau j}\\[2mm]
	h^{R}_{\tau j} \end{array}\right),\qquad
	{\mathcal B}^{\tau}_j=\left(\begin{array}{ccc}
	h^{L}_{\tau j}\\[2mm]
	f^{R}_{\tau j} \end{array}\right).
\label{eq:coupl2}
\end{equation}
In eq.~(\ref{eq:coupl2}),
$f^L_{\tau j}$ and $f^R_{\tau j}$ are defined by eqs.~(\ref{eq:fl}) and (\ref{eq:fr}), respectively, and
\begin{eqnarray}
	h^{L}_{\tau j}&=& (h^{R}_{\tau j})^{\ast}
	=
	m_{\tau}/(\sqrt{2}m_W \cos\beta)
	N_{j3}^{\ast}, 
\end{eqnarray}
with $m_W$ the mass of the $W$ boson, $m_{\tau}$ the mass of the $\tau$-lepton and $N$ the neutralino mixing matrix in the $\tilde{\gamma}, \tilde{Z}, H_1^0, H_2^0$ basis. 
The masses and couplings of the $\tau$-sleptons follow from 
the $\tilde\tau_L - \tilde \tau_R$ mass matrix:
\begin{equation}
{\mathcal{L}}_M^{\tilde \tau}= -(\tilde \tau_R^{\ast},\, \tilde \tau_L^{\ast})
\left(\begin{array}{ccc}
			\ m_{\tilde\tau_R}^2  &
			-m_\tau \Lambda_\tau \\[4mm]
			  -m_\tau \Lambda_\tau 
			& \ m_{\tilde\tau_L}^2  
\end{array}\right)\left(
\begin{array}{ccc}
	\tilde \tau_R\\[4mm]
	\tilde \tau_L \end{array}\right),
\label{eq:mm}
\end{equation}
with $m_{\tilde\tau_R}^2$ and $m_{\tilde\tau_L}^2$ given by eqs.~(\ref{mslr}) and (\ref{msll}) replacing $m_\ell^2$ by $m_\tau^2 $, and 
\begin{eqnarray}
	\Lambda_\tau & = & A_{\tau}-\mu\tan\beta, \label{eq:mlr}
\end{eqnarray}
where $A_{\tau}$ is the trilinear scalar coupling parameter.
The $\tilde \tau$ mass eigenstates are
$(\tilde\tau_1, \tilde \tau_2)=(\tilde \tau_R, \tilde \tau_L) {{\mathcal R}^{\tilde \tau}}^{T}$ with the stau mixing matrix
 \begin{equation}
{\mathcal R}^{\tilde \tau}
         =\left( \begin{array}{ccc}
	\cos\theta_{\tilde \tau} &
	\sin\theta_{\tilde \tau}\\[4mm]
	-\sin\theta_{\tilde \tau} &
	\cos\theta_{\tilde \tau}
\end{array}\right),
\label{eq:rtau}
\end{equation}
and
\begin{equation}
	\cos\theta_{\tilde \tau}=\frac{m_{\tilde\tau_L}^2 -m_{\tilde \tau_1}^2}
	{\sqrt{m_\tau^2\Lambda_\tau^2+(m_{\tilde \tau_1}^2-m_{\tilde\tau_L}^2)^2}}
		,\quad
	\sin\theta_{\tilde \tau}=
	   \frac{m_\tau\Lambda_\tau}
		{\sqrt{m_\tau^2\Lambda_\tau^2+
	(m_{\tilde \tau_1}^2- m_{\tilde\tau_L}^2)^2}        }.
\label{eq:thtau}
\end{equation}
The mass eigenvalues are
\begin{equation}
	 m_{\tilde \tau_{1,2}}^2 = 
		\frac{1}{2}\left(
			(m_{\tilde\tau_L}^2 + m_{\tilde\tau_R}^2)\mp
		\sqrt{
			(m_{\tilde\tau_L}^2-m_{\tilde\tau_R}^2)^2
			+4 m_\tau^2\Lambda_\tau^2
	}
		\right).
\label{eq:m12}
\end{equation}

\section{Density matrix}
\setcounter{equation}{0}
\subsection{Polarization vectors}
\label{spinvectors}
The polarization vectors $s^a_\mu(\neu_j)$ ($a=1,2,3$) of the neutralino $\neu_j$ in the CMS
are chosen so that $\vec{s}_{\chi_j}^{\,1}$ and $\vec{s}_{\chi_j}^{\,2}$ are perpendicular to the momentum of the neutralino $\vec{p}_{\chi_j}$ and $\vec{s}_{\chi_j}^{\,3}$ is parallel to $\vec{p}_{\chi_j}$.
In the reference frame where the four-momentum of the neutralino is given by
\begin{equation}
	p_{\chi_j}^\mu=(E_{\chi_j};0,0,|\vec{p}_{\chi_j}|),
\end{equation}
and where the normal to the production plane is 
\begin{equation}
	\frac{\vec{p}_{\mu^-}\times \vec{p}_{\chi_j}}{|\vec{p}_{\mu^-}\times \vec{p}_{\chi_j}|} = (0,1,0),
\end{equation}
we define the polarization vectors 
\begin{equation}
	s_{\chi_j}^{1\mu}= (0;1,0,0), \qquad	s_{\chi_j}^{2\mu}= (0;0,1,0), \qquad	s_{\chi_j}^{3\mu}= \frac{1}{m_{\chi_j}}(|\vec{p}_{\chi_j}|;0,0,E_{\chi_j}).
\end{equation}

\subsection{Production density matrix: $H$ and $A$ exchange}
\label{proddens}
The interaction Lagrangians are given in eqs.~(\ref{mumuphi}) and (\ref{chichiphi}).
The contribution of the $CP$-even $H$ and $CP$-odd $A$ exchange channels to the 
 production density matrix 
coefficients $P$ and $\Sigma_P^3$, defined by eq.~(\ref{rhoP}), are
\label{proddens.cpc}
 \begin{eqnarray}
	    P_{{R}}   
	        &=&  \frac{1}{2}(2-\delta_{ij}) (1 + P_+^L  P_-^L) \nonumber
\\[1mm]& & 
		g^4[
			  |c_{ij}^{(H)}|^2 |c^{(H\mu)}|^2 |\Delta(H)|^2 s\,s_{ij}^+
			+ |c_{ij}^{(A)}|^2 |c^{(A\mu)}|^2 |\Delta(A)|^2 s\,s_{ij}^- ].
\label{pcpcons}
\\[2mm]
	\Sigma_{{R}}^3 & = & 
		(2-\delta_{ij})(P_+^L + P_-^L) 
\nonumber\\[1mm]
		& & g^4 
		\mbox{Im}(c_{ij}^{(H)} c_{ij}^{(A)\ast})\mbox{Im}( c^{(H\mu)} c^{(A\mu)\ast})
		\mbox{Re} \{ (\Delta(H))(\Delta(A))^*\} s\sqrt{\lambda_{ij}}\eta_j.
\label{sigmacpcons}
\end{eqnarray}
Here $\Delta(H)$, $\Delta(A)$ are the Breit-Wigner propagators defined in eq.~(\ref{hpropagators}), the functions $s_{ij}^\pm$ are defined in eq.~(\ref{s_pm}) and $\lambda_{ij}=s^+_{ij}s^-_{ij}$ is the triangle function.
 
\subsection{Neutralino decay density matrix}
\label{decdens}
The interaction Lagrangians are given in eqs.~(\ref{eq:Lagneutoellsel}) and (\ref{eq:Lagneutotaustau}).
The neutralino decay density matrix, eq.~(\ref{rhoD}),
is 
   \begin{eqnarray} \nonumber
                \rho_{D}^{\lambda_j' \lambda_j} & = & 
           \delta_{\lambda_j' \lambda_j} D +
                          \sum_a \sigma^a_{\lambda_j' \lambda_j}\Sigma^a_{D}.
    \end{eqnarray}
The expansion coefficients $D$ and $\Sigma^a_{D}$ for
the two-body decay into a positive charged lepton of the first two generations
and a right or left slepton are \cite{Bartl:2003tr}:
   \begin{eqnarray}
      D & = & \frac{g^2}{2} |f^{n}_{\ell j}|^2 (m_{\chi_j}^2 -m_{\tilde{\ell}}^2 ),
   \label{DR}
\\
      \Sigma^a_{D} &=&\etarl   g^2 |f^{n}_{\ell j}|^2 m_{\chi_j} (s^a_{\chi_j}
                        \cdot p_{\ell})\eta_j,\quad n=R,L,
   \label{SigmaD}
   \end{eqnarray}
respectively, with $s^a_{\chi_j}$ the neutralino spin-vector defined in Section \ref{spinvectors}, 
$p_{\ell}$ the lepton four-momentum, 
and $\eta_\ell^R=1$ and $\eta_\ell^L=-1$. 
The couplings $f^{n}_{\ell j}$ are defined in eqs.~(\ref{eq:fl}) and (\ref{eq:fr}).

In the CMS 
  \begin{eqnarray}
	\Sigma^3_{D} & = &  -\etarl g^2 |f^{n}_{\ell j}|^2 \frac{m_{\chi_j}^2}{|\vec{p}_{\chi_j}|}
		\left( 
			E_\ell-\frac{	m_{\chi_j}^2 -m_{\tilde{\ell}}^2   }{	2 m_{\chi_j}^2   } E_{\chi_j}	 
		\right)\eta_j, \quad n=R,L.
   \label{SigmaD3}
   \end{eqnarray}

For the decay $\neu_j\to\tau^+\stau^-_{n},\ n=1,2$ 
the coefficients are
   \begin{eqnarray}
                D & = & \frac{g^2}{2} (
                        |a_{nj}^{\tilde \tau}|^2 +|b_{nj}^{\tilde \tau}|^2 )
                                (m_{\chi_j}^2 - m_{\tilde{\tau}}^2 ),
\label{Dtau}\\
                \Sigma^a_{D} &=& - g^2 (
                        |a_{nj}^{\tilde \tau}|^2-|b_{nj}^{\tilde \tau}|^2 )
                        m_{\chi_j}      (s^a_{\chi_j} \cdot p_{\ell})\eta_j.
   \label{SigmaDtau}
   \end{eqnarray}
The couplings $a^{\tilde \tau}_{nj}, n=1,2$ are defined in eq.~(\ref{eq:taucouplings}).

In the CMS 
  \begin{eqnarray}
	\Sigma^3_{D} & = &  g^2  (|a_{nj}^{\tilde \tau}|^2-|b_{nj}^{\tilde \tau}|^2 )              
			\frac{m_{\chi_j}^2}{|\vec{p}_{\chi_j}|}
		\left( 
			E_\ell-\frac{	m_{\chi_j}^2 -m_{\tilde{\tau}}^2   }{	2 m_{\chi_j}^2   } E_{\chi_j}	 
		\right)\eta_j, \quad n=1,2.\ 
  \label{SigmaD3tau}
  \end{eqnarray}


The decay density matrix coefficients for the charge conjugated processes, 
$\neu_j\to\ell^-\slepton_n^+, n=R,L$ and $\neu_j\to\tau^-\stau_n^+, n=1,2$  
are obtained inverting the sign of $\Sigma^a_{D}$ in
eqs.~(\ref{SigmaD}),  (\ref{SigmaD3}), 
(\ref{SigmaDtau}) and (\ref{SigmaD3tau})


\end{appendix}



\begin{thebibliography}{99}

\bibitem{hefreports}Proceedings of {\it Prospective Study of Muon Storage Rings at CERN}, 
Eds. B.~Autin, A.~Blondel, J.~Ellis, CERN yellow report,
CERN 99-02, ECFA 99-197, April 30 (1999);
\\
C.~Bl\"ochinger et al., Higgs working group of the ECFA-CERN study on Neutrino Factory \& Muon Storage Rings at CERN, {\it Physics Opportunities at $\mu^+\mu^-$ Higgs Factories}
CERN-TH/2002-028, [arXiv:hep-ph/0202199];
\\
A.~Blondel {\it et al.},
``ECFA/CERN studies of a European neutrino factory complex,''
CERN-2004-002.

\bibitem{mucolhiggs}
R.~Casalbuoni et al.,  JHEP \textbf{9908} (1999) 011;
\\
V.~Barger, M.S.~Berger, J.F.~Gunion, T.~Han, in {\it Proc. of the
APS/DPF/DPB Summer Study on The Future of Particle Physics (Snowmass
2001)}, ed. R.~Davidson and C.~Quigg, 
[arXiv:hep-ph/0110340];
\\
V.~Barger, M.S.~Berger, J.F.~Gunion, T.~Han, Nucl.\ Phys.\ Proc.\ Suppl.\ \textbf{51A} (1996) 13.

\bibitem{barger1}
V.~Barger, M.S.~Berger, J.F.~Gunion, T.~Han, Phys.\ Rep.\ \textbf{286} (1997) 1.

\bibitem{higgsneutralino}
H.~Fraas, et al.;
in preparation

\bibitem{asakawa}
E.~Asakawa, A.~Sugamoto and I.~Watanabe,
Eur.\ Phys.\ J.\ C {\bf 17} (2000) 279  [arXiv:hep-ph/0004005];\\
E.~Asakawa, S.~Y.~Choi and J.~S.~Lee,
Phys.\ Rev.\ D {\bf 63}, 015012 (2001)
[arXiv:hep-ph/0005118].

\bibitem{gudi.majorana}
G.~Moortgat-Pick and H.~Fraas,
Eur.\ Phys.\ J.\ C {\bf 25} (2002) 189
[arXiv:hep-ph/0204333].

\bibitem{HK} H.~Haber, K.~Kane, 
Phys.\ Rep.\ \textbf{117} (1985) 75.

\bibitem{GH} J.~Gunion, H.~Haber, 
Nucl.\ Phys.\ \textbf{B 272} (1986) 1.

\bibitem{gudi99}
G.~Moortgat-Pick, H.~Fraas, A.~Bartl and W.~Majerotto,
Eur.\ Phys.\ J.\ C {\bf 9} (1999) 521
[Erratum-ibid.\ C {\bf 9} (1999) 549]
[arXiv:hep-ph/9903220].

\bibitem{Bartl:2002bh}
A.~Bartl, K.~Hidaka, T.~Kernreiter and W.~Porod,
Phys.\ Rev.\ D {\bf 66}, 115009 (2002)
[arXiv:hep-ph/0207186].

\bibitem{Haber94}H.~Haber, Proceedings of the 21st SLAC Summer Institute on Particle Physics: Spin Structure in High Energy Processes, SLAC, Stanford, CA 1993. 

\bibitem{Bartl:2003tr}
A.~Bartl, H.~Fraas, O.~Kittel and W.~Majerotto,
Phys.\ Rev.\ D {\bf 69}, 035007 (2004)
[arXiv:hep-ph/0308141].

\bibitem{TDR}
``TESLA Technical Design Report Part III: Physics at an e+e- Linear Collider,''
arXiv:hep-ph/0106315.

\bibitem{Nojiri.Boos.Martyn}
M.~M.~Nojiri,
Phys.\ Rev.\ D {\bf 51} (1995) 6281
[arXiv:hep-ph/9412374];
\\
M.~M.~Nojiri, K.~Fujii and T.~Tsukamoto,
Phys.\ Rev.\ D {\bf 54} (1996) 6756
[arXiv:hep-ph/9606370].
\\
E.~Boos, H.~U.~Martyn, G.~Moortgat-Pick, M.~Sachwitz, A.~Sherstnev and P.~M.~Zerwas,
Eur.\ Phys.\ J.\ C {\bf 30} (2003) 395
[arXiv:hep-ph/0303110];
\\
H.~U.~Martyn,
[arXiv:hep-ph/0406123].

\bibitem{Fraas:2003cx}
H.~Fraas, F.~Franke, G.~Moortgat-Pick, F.~von der Pahlen and A.~Wagner,
Eur.\ Phys.\ J.\ C {\bf 29} (2003) 587
[arXiv:hep-ph/0303044].

\bibitem{poltransv}
D.~Atwood and A.~Soni,
Phys.\ Rev.\ D {\bf 52} (1995) 6271
[arXiv:hep-ph/9505233];\\
B.~Grzadkowski and J.~F.~Gunion,
Phys.\ Lett.\ B {\bf 350} (1995) 218
[arXiv:hep-ph/9501339];\\
J.~F.~Gunion, B.~Grzadkowski and X.~G.~He,
Phys.\ Rev.\ Lett.\  {\bf 77} (1996) 5172
[arXiv:hep-ph/9605326];\\
J.~F.~Gunion and J.~Pliszka,
Phys.\ Lett.\ B {\bf 444} (1998) 136
[arXiv:hep-ph/9809306];\\
B.~Grzadkowski and J.~Pliszka,
Phys.\ Rev.\ D {\bf 60} (1999) 115018
[arXiv:hep-ph/9907206].

\bibitem{Hall:zn}
L.~J.~Hall and J.~Polchinski,
Phys.\ Lett.\ B {\bf 152}, 335 (1985).

\bibitem{HDECAY} A.~Djouadi, J.~Kalinowski, M.~Spira,
Comput.\ \ Phys.\ \ Commun.\ \ \textbf{108} (1998) 56.

\bibitem{Allanach:2002nj}
B.~C.~Allanach {\it et al.},
in {\it Proc. of the APS/DPF/DPB Summer Study on the Future of Particle Physics (Snowmass 2001) } ed. N.~Graf,
Eur.\ Phys.\ J.\ C {\bf 25} (2002) 113
[eConf {\bf C010630} (2001) P125]
[arXiv:hep-ph/0202233].

\end{thebibliography}
\end{document}